# MediTools – Medical Education Powered by LLMs

Amr Alshatnawi[1*], Remi Sampaleanu[1*], David Liebovitz, MD[2]

[1]*The University of Chicago, Chicago, IL*

[2]*Northwestern University, Chicago, IL*

**These authors contributed equally*

---

**Abstract**

**Background**

Artificial Intelligence (AI) has been advancing rapidly and with the advent of large language models (LLMs) in late 2022, numerous opportunities have emerged for adopting this technology across various domains, including medicine. These innovations hold immense potential to revolutionize and modernize medical education.

**Methods**

Our research project involved creating *MediTools – AI Medical Education*, a prototype application built using Streamlit and coded in Python, leveraging various APIs and LLMs. We conducted a survey using Qualtrics, gathering feedback from 10 medical professionals and students. Data analysis was performed using Python.

**Results**

We successfully developed *MediTools*, which includes three main tools: a Dermatology Case Simulation tool, an AI-enhanced PubMed tool, and a Google News tool. The Dermatology Case Simulation tool uses real patient images depicting various dermatological conditions and enables interaction with LLMs acting as virtual patients. The AI-enhanced PubMed tool facilitates engaging with LLMs to gain deeper insights into research papers, while the Google News tool provides LLM-generated summaries of medical articles. The survey gathered preliminary feedback, highlighting the application's effectiveness in providing realistic LLM-powered tools that could enhance learning outcomes. Participants expressed high satisfaction with *MediTools* and believed that the integration of AI and LLMs in medical education could lead to improved learning outcomes. These insights are valuable for the further development and refinement of the application.

**Conclusion**

This research underscores the potential of AI-driven tools in transforming and revolutionizing medical education, offering a scalable and interactive platform for continuous learning and skill development. It illustrates the value of integrating AI technologies into educational frameworks.

## Introduction

When ChatGPT was released in late 2022, not even the OpenAI creators themselves anticipated the level of interest and use that would shortly follow. In the years since, artificial intelligence (AI) and large language models (LLMs) like ChatGPT have seen widespread adoption and application, with nearly every single industry trying to take advantage of this new technology in some form or another. Unfortunately, when it comes to new applications and technological advances, the healthcare industry is often slow to follow. With many laws and regulations in place to ensure patient confidentiality and safety, it can be a difficult and lengthy process for new healthcare applications to be approved (Abd-alrazaq, et al., 2023). There are, however, several domains within healthcare that could employ these new technologies with little direct patient risk. Medical education is one such domain that will likely be revolutionized by generative AI and LLMs in the coming decades. These technologies will undoubtedly serve as the groundwork for new learning tools and applications that can help students practice crucial clinical skills, whether that be by producing case studies, acting as virtual patients, or by helping to guide them through diagnosis workflows. The use of LLMs in medical education is inevitable and we need to learn how to best utilize this technology while also recognizing its limitations (Furfaro, et al., 2024).

LLMs present numerous opportunities in medical education, holding significant potential to modernize medical education and revolutionize the learning process (Li & Li, 2023). As we move forward, it is crucial to explore ways to integrate this technology into real-world use cases effectively. This includes assessing the effectiveness of incorporating LLMs in clinical training and evaluating their potential to address workflow challenges in medicine. This profound technology offers potential enhancement in learning, diagnostic simulations, and access to medical knowledge. This highlights an opportunity in medical education that calls for more research, particularly studies that explore the use of AI applications to enhance and improve the learning experience (Nagi, et al., 2023; Li & Li 2023).

In response to this identified opportunity, we have developed a prototype application called *MediTools – AI Medical Education* in this research project. In collaboration with experts from Northwestern Medicine, we created an online hub of tools for medical students and professionals. Our application allows individuals to practice obtaining medical histories, practice their diagnostic workflows, and to stay up to date with the latest news and research articles in their fields of interest. Our first tool is a dermatology case simulation tool, where we present users with real patient images that represent various skin conditions, allowing them to interact with an LLM acting as a virtual patient using text, and voice features. This tool incorporates various features that allow the user to get feedback and practice their diagnostic skills and clinical decision-making abilities. We have also created two additional LLM-powered tools: An AI-Enhanced PubMed and Google News tool that allows users a modern way to access medical literature and news. By harnessing the ability of LLMs to accurately simulate real patient interactions and output results in interpretable natural English, we are able to provide educational tools that use less manpower, time, and effort than traditional methods.

## Methods

### I. Application Software & Codebase

At its core, *MediTools* is built using Python and Streamlit, an open-source framework used for application development and deployment. CSS and HTML scripts are also used extensively throughout the application to display and modify the user interface. For reproducibility, all application scripts and documents are stored in a GitHub repository (https://github.com/NM-Streamlit-Team/meditools). This repository was used to manage the application codebase, and track ongoing issues, errors, and new features. It also includes all application documentation and specifies the packages and libraries required to develop the application. *MediTools* was built in a standard Linux environment using WSL (Windows Subsystem for Linux). The user-facing component of the application is structured simply, with a home page, and separate pages for each individual tool that are accessible from a sidebar in the application. User navigation across pages and tool interfaces is managed using Streamlit's session state variables stored in a dictionary-like structure. This approach further enables the storage of important variables like passwords and chat histories, ensuring they persist even when switching between different tools. Additionally, several backend files contain the application's functions and LLM prompts. To handle the feedback form and email communications between users and the *MediTools* team on our home page, we use Twilio & SendGrid.

### II. Technology Stack for Tools

Both the Dermatology Case Simulation Tool & the AI-enhanced PubMed tool are powered by LLMs that allow users to chat and interact with it. This LLM component was built using LangChain, a framework for building applications with LLMs. We used a combination of LangChain/Streamlit packages to help us create LLM chains that can handle chat memory and format prompt templates. This framework is utilized throughout our application to integrate LLM capabilities into our tools. For example, we use this framework in the Dermatology case simulation tool to generate synthetic lab tests using OpenAI's GPT-4o model and we use it to create feedback reports. Each of these features has its own LLM prompt to achieve the desired results. To provide features such as Text-to-Speech (TTS) and Speech-to-Text (STT), we use OpenAI's TTS API and their STT API using the whisper-1 model. We give users the ability to select different LLMs (OpenAI, Anthropic, Meta, etc.) for their interaction. Based on the selected model, the respective API belonging to that LLM provider is called. For OpenAI models, we use OpenAI's API, while for all other models, we use the OpenRouter service API.

For patient images in the Dermatology case Simulation tool, we used a Kaggle dataset that contains images of different dermatological conditions taken from a public portal called DermNet. For this initial prototype application, these images are stored on GitHub in a directory structure organized by condition, with files named according to their type. This setup allows for easy access and retrieval of images using a simple function that selects a random image. Based on the directory that the image is stored in and the file name, we pass the condition to the LLM in a prompt and create a case for the user.

For the AI-enhanced PubMed tool, we use multiple PubMed E-utilities APIs. Based on the user's entered search query, we pass this query to a PubMed API that retrieves PubMed

Identifiers (PMID) that match the query. We then pass this list of PMIDs to another API that retrieves metadata about the article in XML format. We use a custom function that we developed to help us extract metadata from the XML data like title, authors, year, abstract, and more. We also check if that article has a PubMed Central Identifier (PMCID), which indicates if this article has been archived on PubMed Central. PubMed Central is an index of full-text papers that can be accessed using a standard URL with a specific PMCID. If the retrieved article has a PMCID, users then can select this paper and initiate an LLM chat to interact with it about the content of the paper. The full text of the paper is retrieved using a service called Diffbot, which employs advanced machine learning and computer vision technologies to extract data from web pages. This text is then passed in a prompt to the LLM.

The Google News tool has a very simple setup. We structure a search query based on the user's selected specialties and configure a search function with the chosen tool parameters using Google Serper, which is a very fast and effective Google search API. The search query is then passed to the API to generate results. For generating article summaries, we use OpenAI's GPT-4o model and a LangChain LLM summary chain that allows us to produce summaries using only the article's URL.

## III. User Survey and Feedback Evaluation

We developed a survey to gather initial feedback on the effectiveness and user satisfaction of *MediTools*. This data collection was informal and has not been reviewed by an Institutional Review Board (IRB). It involved a brief usability survey with 10 healthcare professionals to gather preliminary feedback on the application. No identifiable or sensitive information was collected, and participation was entirely voluntary. The survey is administered online using Qualtrics.

We reached out to 10 participants, consisting of various healthcare professionals and students, either by email or in-person, and obtained consent from all participants. Participants were selected using convenience sampling, which involved choosing individuals who were easily accessible to us. The survey consists of 25 questions divided into 5 sections: demographics, dermatology case simulation tool feedback, AI-enhanced PubMed tool feedback, Google News feedback, and overall feedback with a mix of multiple-choice questions and short answers.

Data analysis was conducted using Python in an anaconda Jupyter notebook. We utilized various libraries such as Pandas for data manipulation, NumPy for numerical operations, and Matplotlib/Seaborn for data visualization. The analysis included descriptive statistics to summarize the responses, and visualizations to represent the distributions and trends in the feedback. This approach allowed us to efficiently handle the survey data and extract meaningful insights into the effectiveness and user satisfaction of *MediTools*.

## IV. Workflow of Implemented Tools

To clarify the end-end design for implemented tools, **Figures 1 & 2** summarize the workflow of the Dermatology Case Simulation Tool and the AI-Enhanced PubMed Tool developed in this study. These visuals highlight the major system components, data flow, and interaction between user input, orchestration logic, external, services, and response generation.

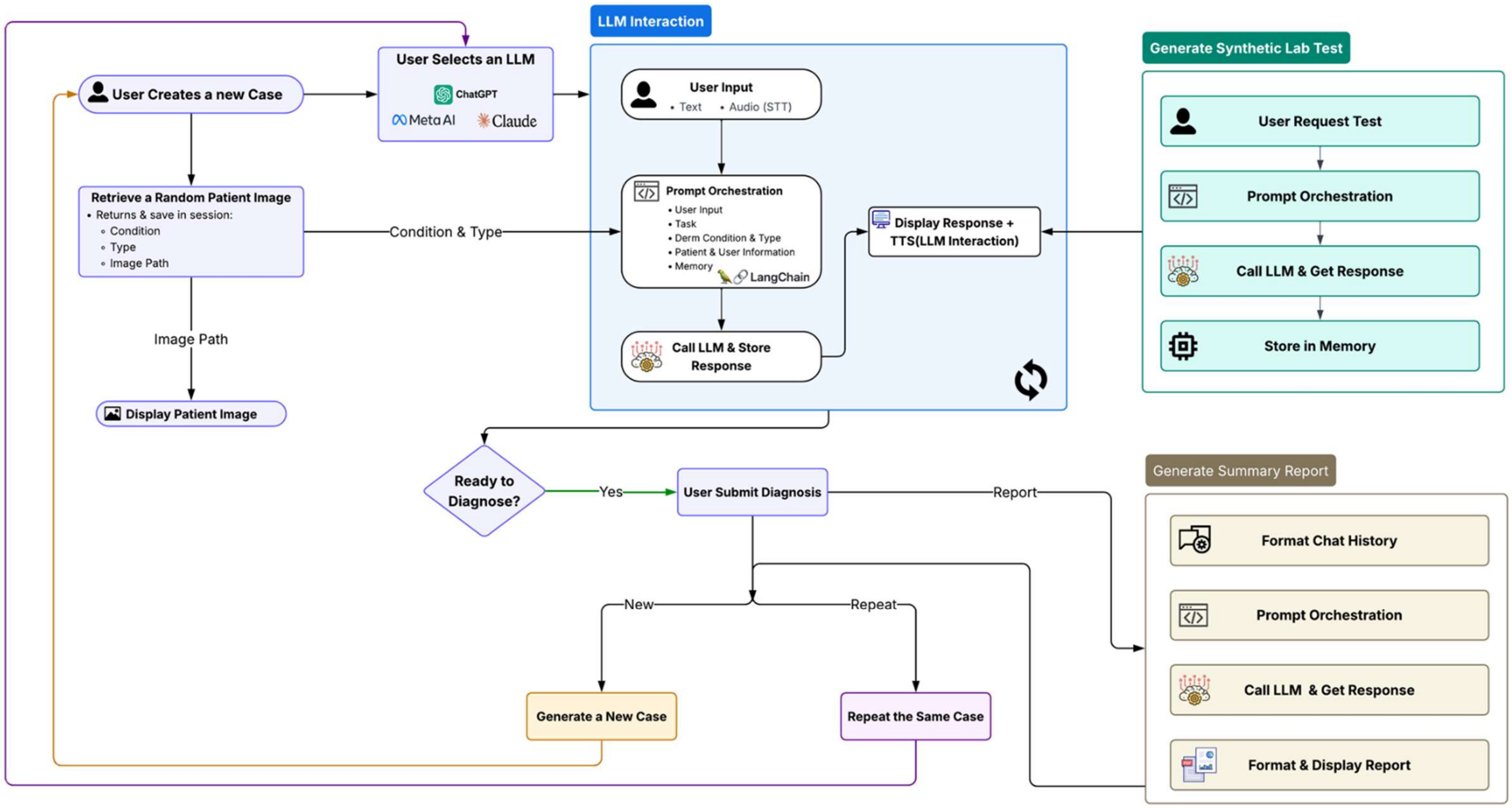

**Figure 1.** High-level workflow of the Dermatology Case Simulation Tool, showing case generation, patient image retrieval, LLM interaction, diagnosis submission, and report generation.

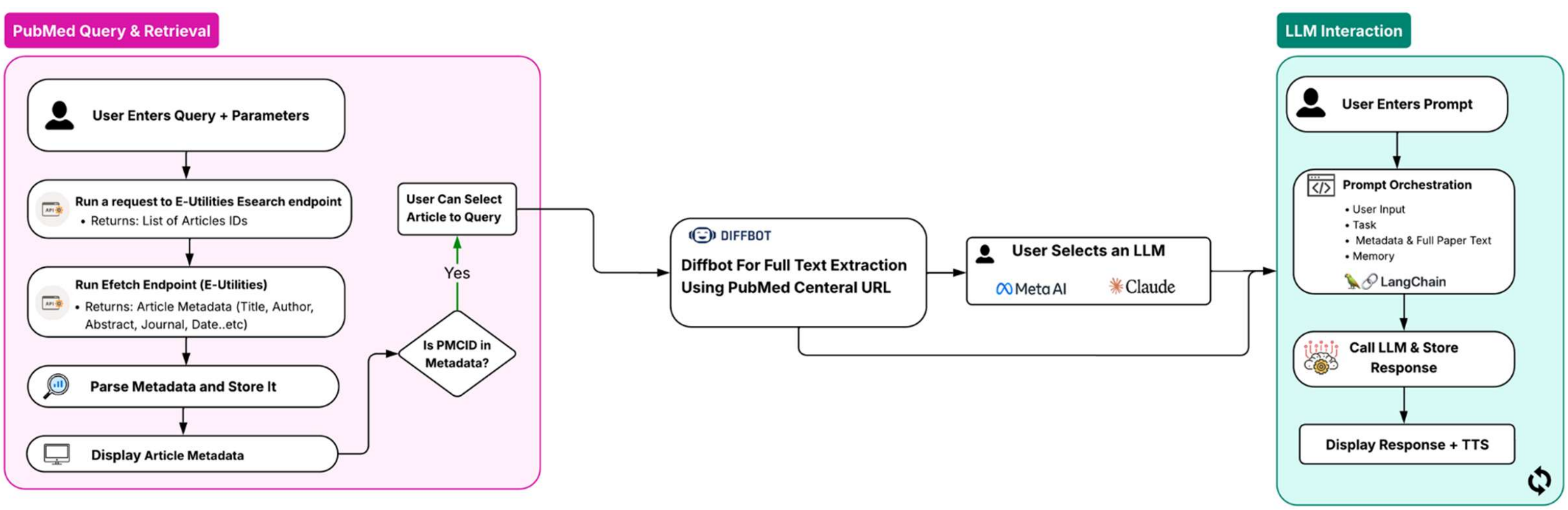

**Figure 2.** High-level workflow of the AI-Enhanced PubMed Tool, showing article retrieval, metadata extraction, full-text processing, LLM interaction, and response generation.

Together, these workflows provide a high-level view of how the tools were structured and executed in practice. The following Results section presents the implemented tool from the user perspective including their interface, functionality, and representative outputs.

## Results

### A. Dermatology Case Simulation Tool

#### A.1 Tool Overview

The dermatology case simulation tool harnesses the power of LLMs to provide a simulated clinic environment in which students and professionals can practice realistic patient interactions and their diagnosis workflows. Upon launching the tool, users are first asked to provide their name, and to click a button which generates their first case (see **A.3 Case Creation**). Users are then prompted for their model preferences (e.g. ChatGPT, llama, claude), and their desired feedback configuration (**Figure 3)**. Once all parameters have been set, users can begin their virtual patient interaction via text chat or with their voice. At any time during their interaction, users are able to click a button to view an image of their patient's condition, generate a report summarizing their conversation, or submit a guess as to the patient's condition (**Figure 4**). Users are then able to obtain feedback for their previous interaction, repeat the case, or try a different case (see section **A.4** for more details). Each case is unique, and consists of a patient with a new name, personality, and dermatological condition. By providing the condition to the model behind the scenes, the LLM is able to construct a realistic and convincing scenario using information about the condition found in its database.

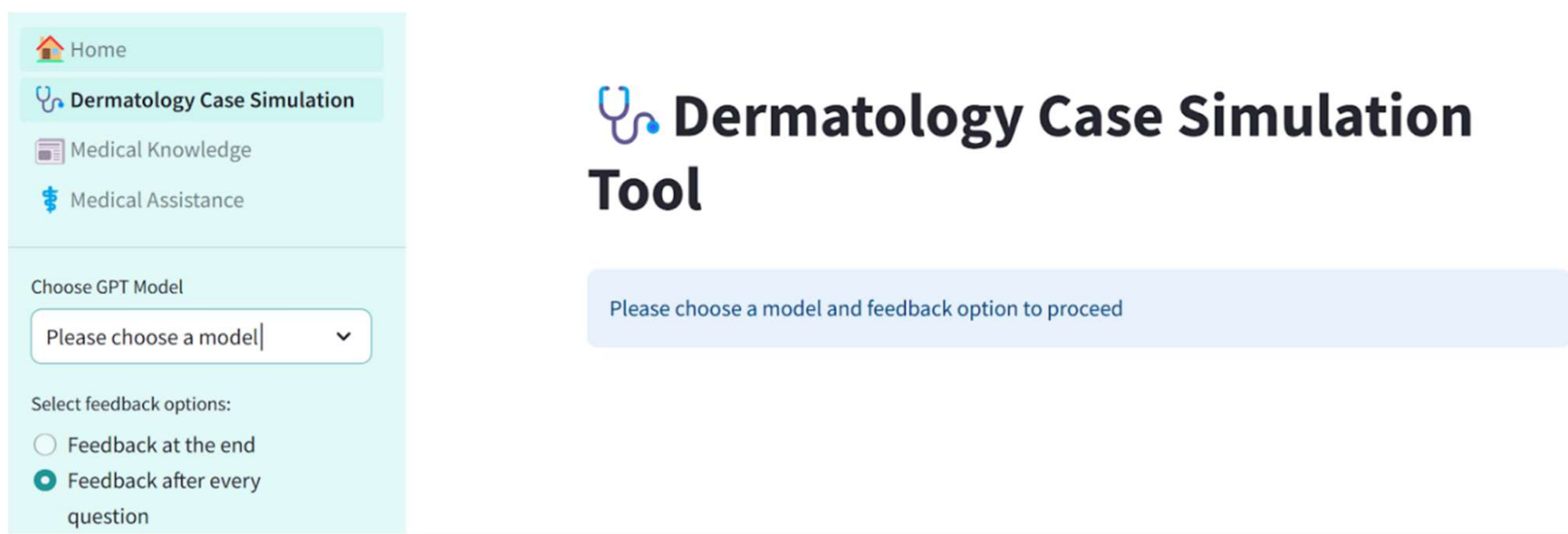

**Figure 3.** Case simulation tool prior to selecting LLM model preference. Users are only able to proceed with the interaction once a model has been selected.

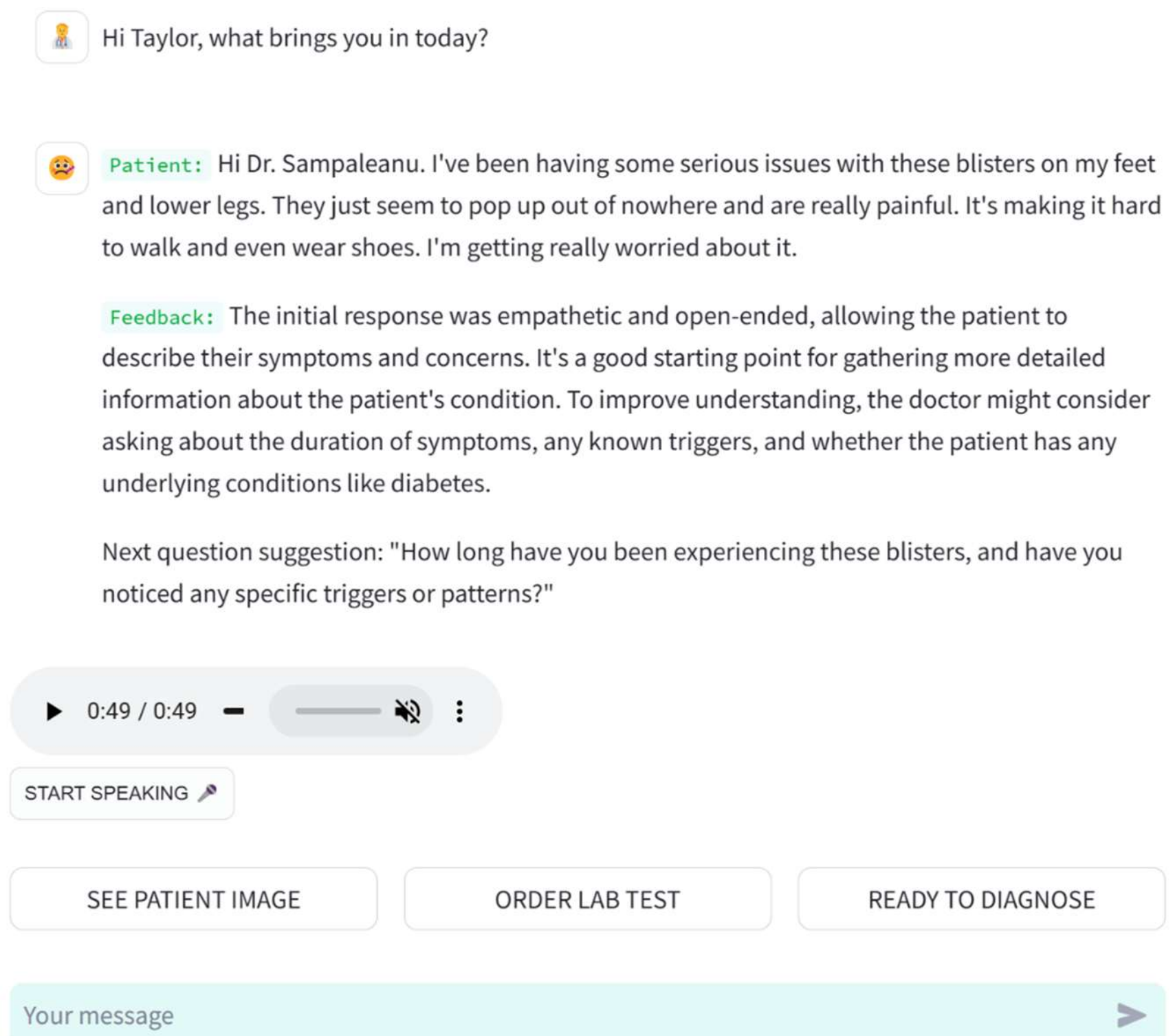

**Figure 4.** Sample interaction (with feedback after every question selected as the feedback option) showing the three buttons users can click to see the condition image, order a lab test, or submit their guess.

### A.2 Feedback

Users are able to select one of two different feedback configurations from the sidebar: 'Feedback at the end' and 'Feedback after every question' (see **Figure 3**). When the former is selected, no live feedback or advice is given to the user during their interaction with the virtual patient. The latter configuration enables live feedback and will gently remind the user of proper procedures and improvements in their communication style whenever it deems a user response to be inappropriate or inadequate (**Figure 4**). This evaluation of user input is specified within the virtual patient model prompt. Regardless of the configuration chosen, at any time after submitting a condition guess, users can generate a summary report for their interaction. This report has three main components: 1) information about the condition presented by the patient (symptoms, treatment, etc.), 2) a transcript of the user's interaction with the patient, and 3) performance feedback for the interaction, including things the user did well and things that could be improved.

### A.3 Case Creation, Images, & Lab Test

Whenever the create new case button is clicked, several actions occur in sequence behind the scenes. First, all session state variables pertaining to the previous case are reset. This includes the stored patient name, condition image, condition name and type, and patient personality. The new name and personality type is randomly selected from a list, while the condition name and images are directly retrieved from a random path name in our image database. These new values are then stored in their respective session state variables, allowing them to be easily accessed no matter the page state. Lab tests are generated using a separate LLM chain, which is instructed to provide reasonable test results of the type requested by the user. The lab results are then formatted and injected into the ongoing chat-stream, to be viewed and interpreted by the user (**Figures 5, 6**).

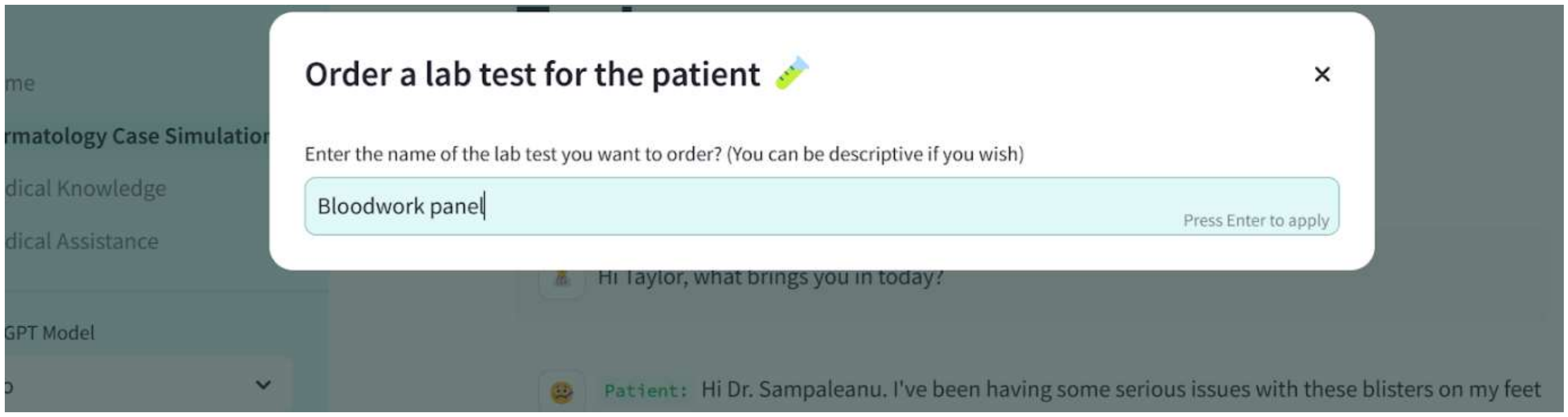

**Figure 5.** Dialog box for submitting a request for lab tests.

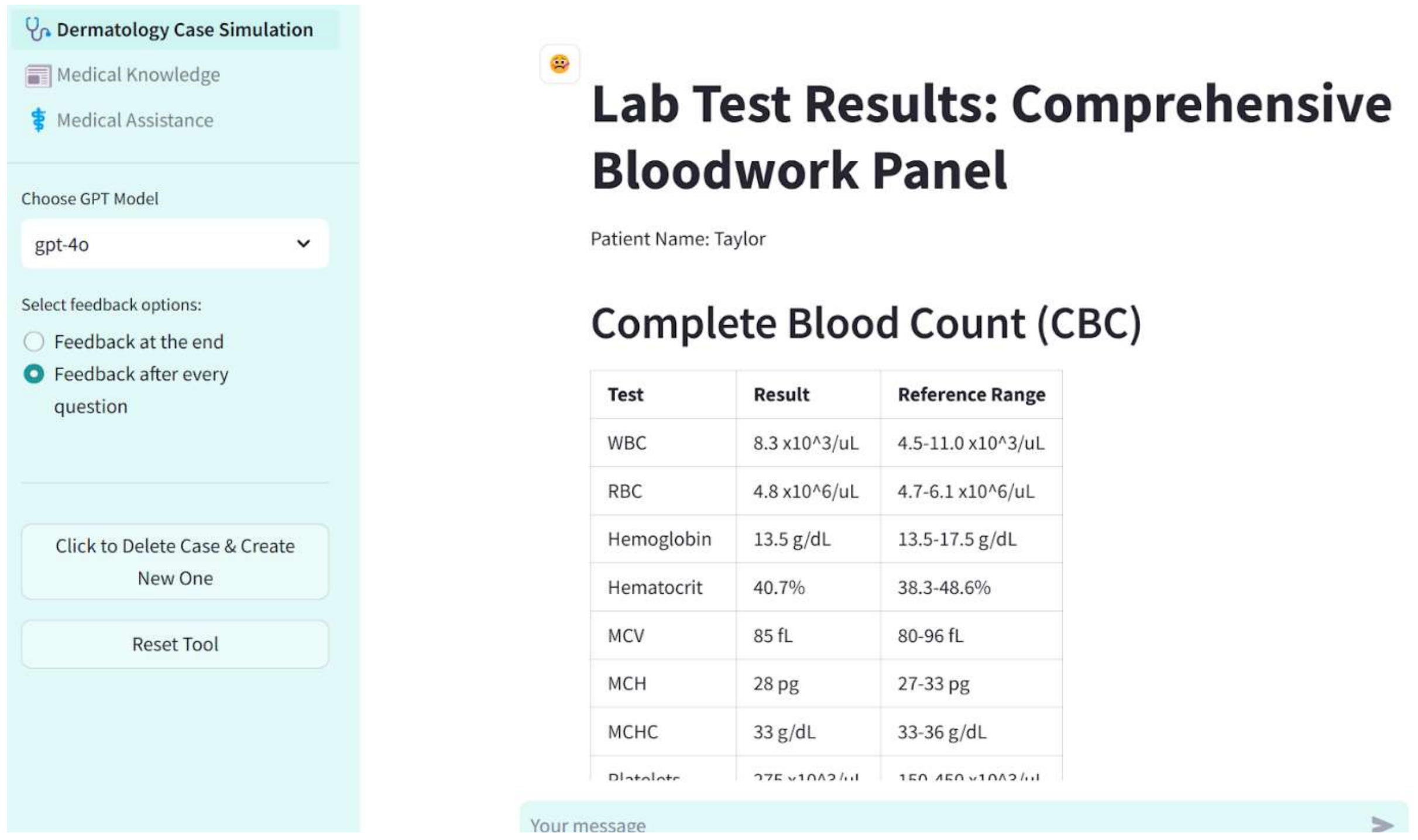

**Figure 6.** Sample lab test results for a patient with Bullous disease, generated via bloodwork request.

### A.4 Submitting a Guess

The guess submission process is relatively simple. Upon clicking the ready to diagnose button, users are prompted to type their guess into a popup dialog window. *MediTools* then uses approximate string matching (also known as 'fuzzy' string matching) to compare the user's input with the real patient condition. Both of these values are strings, and can thus be compared using Levenshtein distance to determine true matches. We chose to use SeatGeek's thefuzz package for this task, which contains several different ratio calculation functions that are well suited to this task. We chose to calculate our match ratio using the token set ratio of the two strings, which is ideal when comparing two strings where word order does not matter. After some experimentation, we selected a cutoff of 0.7 as the threshold for matches. Once users have submitted their guess, they are greeted by a message stating whether or not they were correct, and the full condition name. Users can then repeat the previous scenario, try a new one, or view their last interaction's summary report (**Figure 7**).

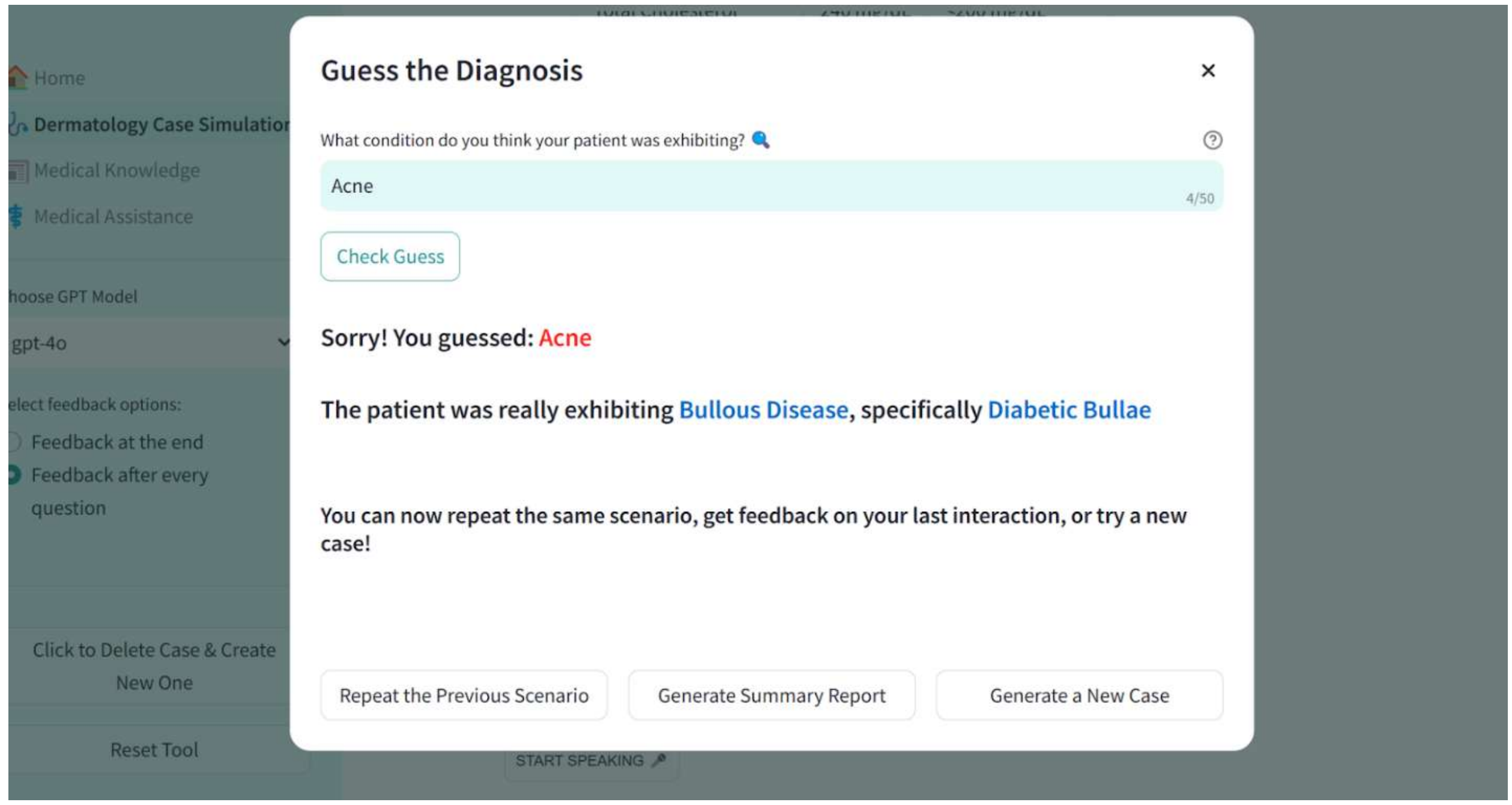

**Figure 7.** Post-guess screen showing the options available to the user after submitting their diagnosis. Users can repeat the exact same scenario, generate a summary report of their interaction, or try a new case.

## B. Google News Tool

### B.1 Tool Overview

The Google News tool tab of the medical knowledge tool is very straightforward and intuitive to use. It allows users to rapidly browse an array of news articles in their fields of interest, view AI-generated summaries of these articles, and access the original articles if desired. We envision this tool as a starting point in a student or professional's routine, keeping them updated with current events and advancements in the domain they are pursuing. To generate their personalized news results, prior to clicking the 'Generate Results' button, users are able to select various filters and parameters to enhance their query. Users can select the recency of their results, add keywords to their search, and pick the number of article summaries generated (**Figure 8**). This last number is divided equally by the selected fields of interest when generating output. For example, if the user indicates they would like to generate seven article summaries, the tool might generate two dermatology article summaries, two cardiology summaries, and three rheumatology summaries (assuming those fields of interest have been selected by the user). If at any time the Google Serper API is not able to retrieve a sufficient number of articles using the selected filters, the user is notified by a warning message stating the field of interest that is lacking in available sources (once all other available article summaries have been generated).

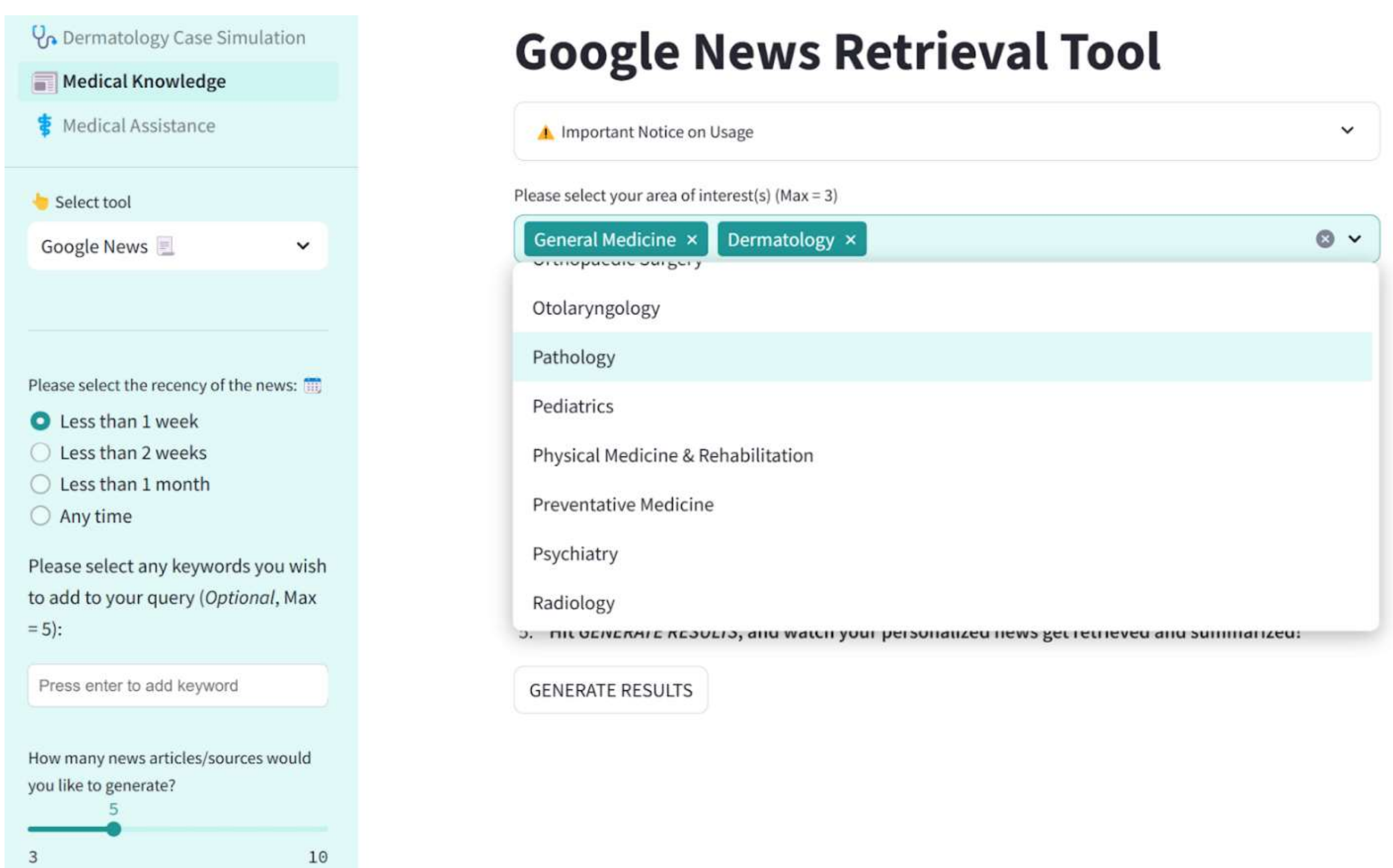

**Figure 8.** Parameters/filters available for user configuration, allowing for personalized results.

## B.2 Article Retrieval

When a request for articles is made by the user, the application first constructs the search query. This is done by combining any selected keywords, the selected topic of interest, and language specifying a search for advancements, updates, etc., into a string. This string is then passed to Google Serper, a fast and effective google news search API. The results from this query are held within a dictionary. This dictionary contains key-value pairs for relevant fields like the article URL or publishing date. Assuming the articles have passed the recency filter selected by the user, the URLs are then passed to Langchain's UnstructuredURLLoader function, which extracts a string representation of the webpage's data. An LLM model, specifically OpenAI's 'gpt-4o', is used to summarize the webpage data and create the article summaries ultimately outputted to the user's screen (see **Figure 9** for sample output).

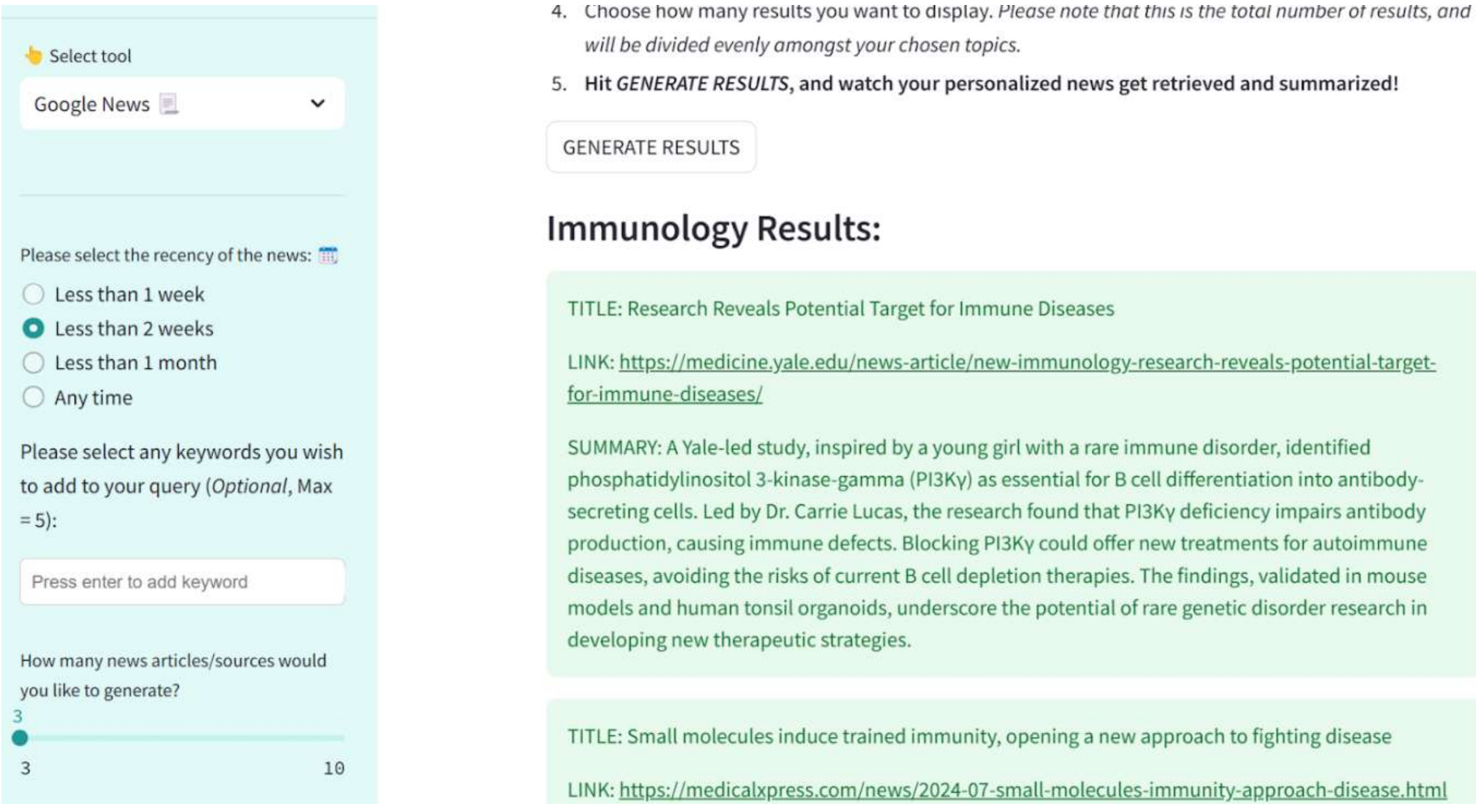

**Figure 9.** Sample output, showing immunology news and advancements within the last two weeks.

## C. AI-Enhanced PubMed Tool

### C.1 Tool Overview & Home Page

The AI-enhanced PubMed Tool enables users to select research papers and query their content using Large Language Models (LLMs). This tool offers a quick and efficient way to understand and gain insights from medical literature with the assistance of advanced AI capabilities. **Figure 10** below illustrates the home page of this tool, where users can enter their search queries in the search field and adjust search parameters using the sidebar. Users can configure how many papers to retrieve and the recency of the articles.

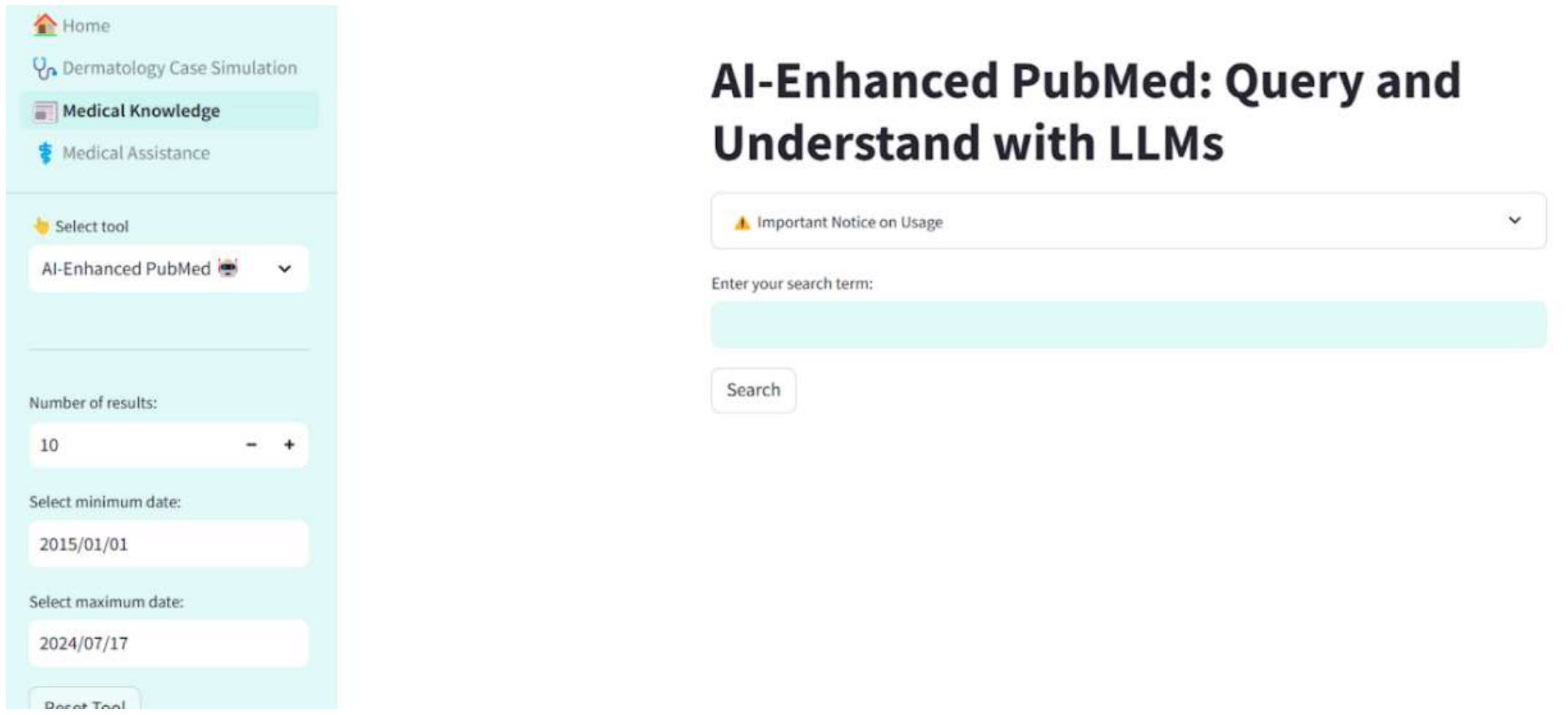

**Figure 10.** Home page of AI-enhanced PubMed Tool, displaying the sidebar options and the search submitting field.

### C.2 Retrieved Articles & Selection Option

After the user enters a search query, a list of matching articles is displayed in block format. Each block provides relevant information about the paper, including the title, PubMed ID (PMID), authors, published year, journal, link to the paper on PubMed, and the abstract. If the paper is archived on PubMed Central, a button will appear at the bottom of the block that allows users to select that paper to query with an LLM. This is illustrated in **Figure 11** below.

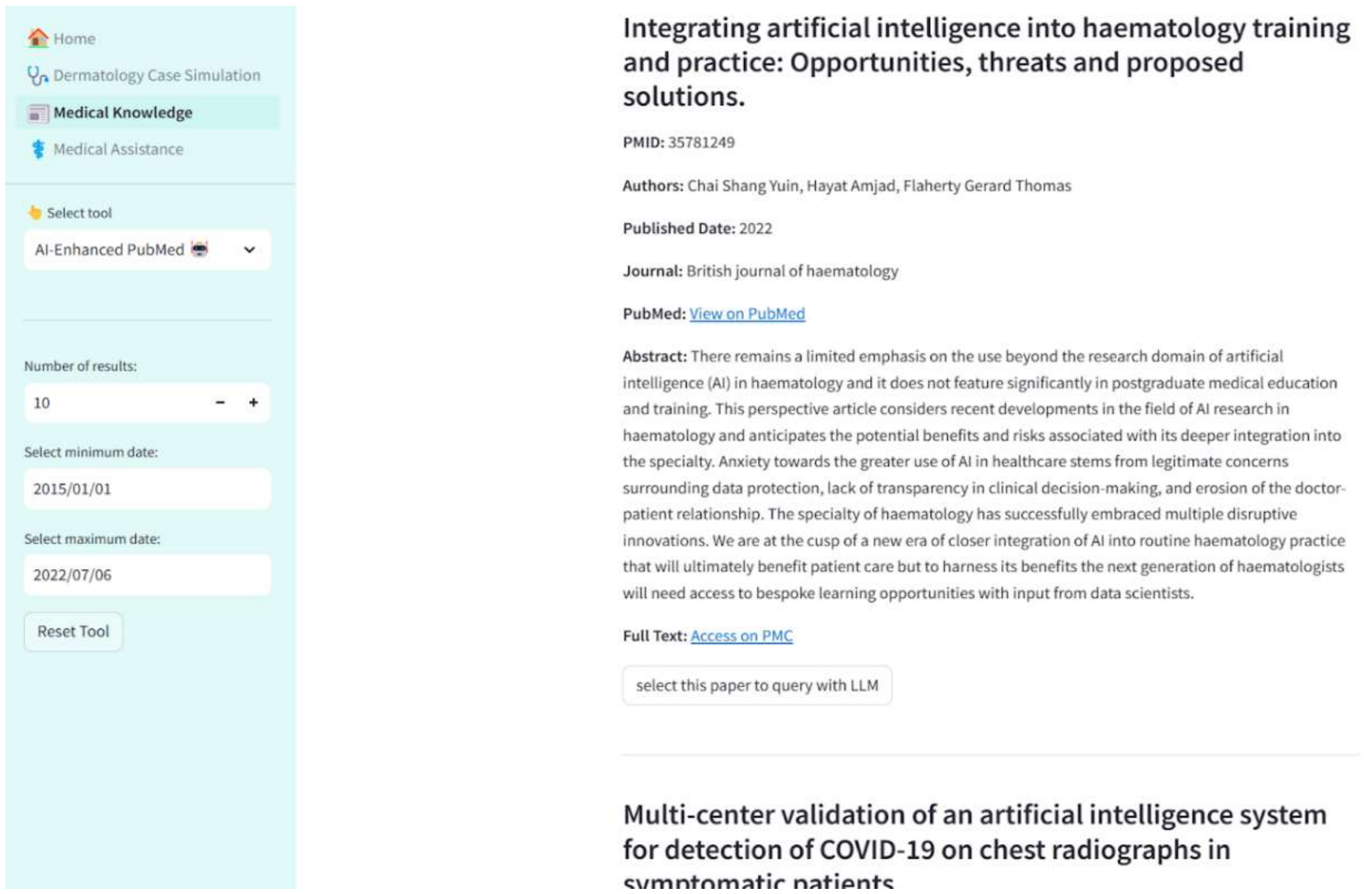

**Figure 11.** Retrieved articles and the option to select an article in the AI-enhanced PubMed Tool.

### C.3 LLM Selection & Chat Interface

When an article is selected, the full text is retrieved, and the user is given the option to choose an LLM model to continue as shown in **Figure 12.**

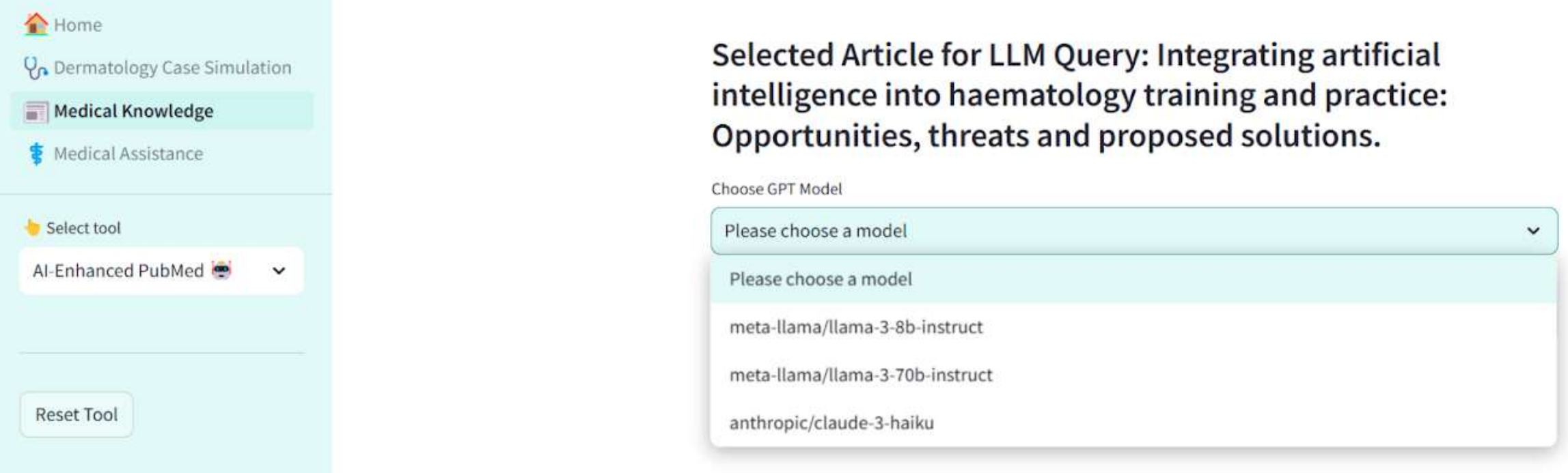

**Figure 12.** LLM selection interface

When the LLM is selected, a chat interface is displayed. The full retrieved text of the paper is passed to the model along with any user query using a prompt. **Figure 13** presents this interface. Links to a PDF version of the paper and the full text on PubMed Central are provided to the user in the sidebar, giving them access to the paper while allowing them to ask questions about the content. The user can prompt the LLM with any questions to help understand aspects of the paper such as the methods or results.

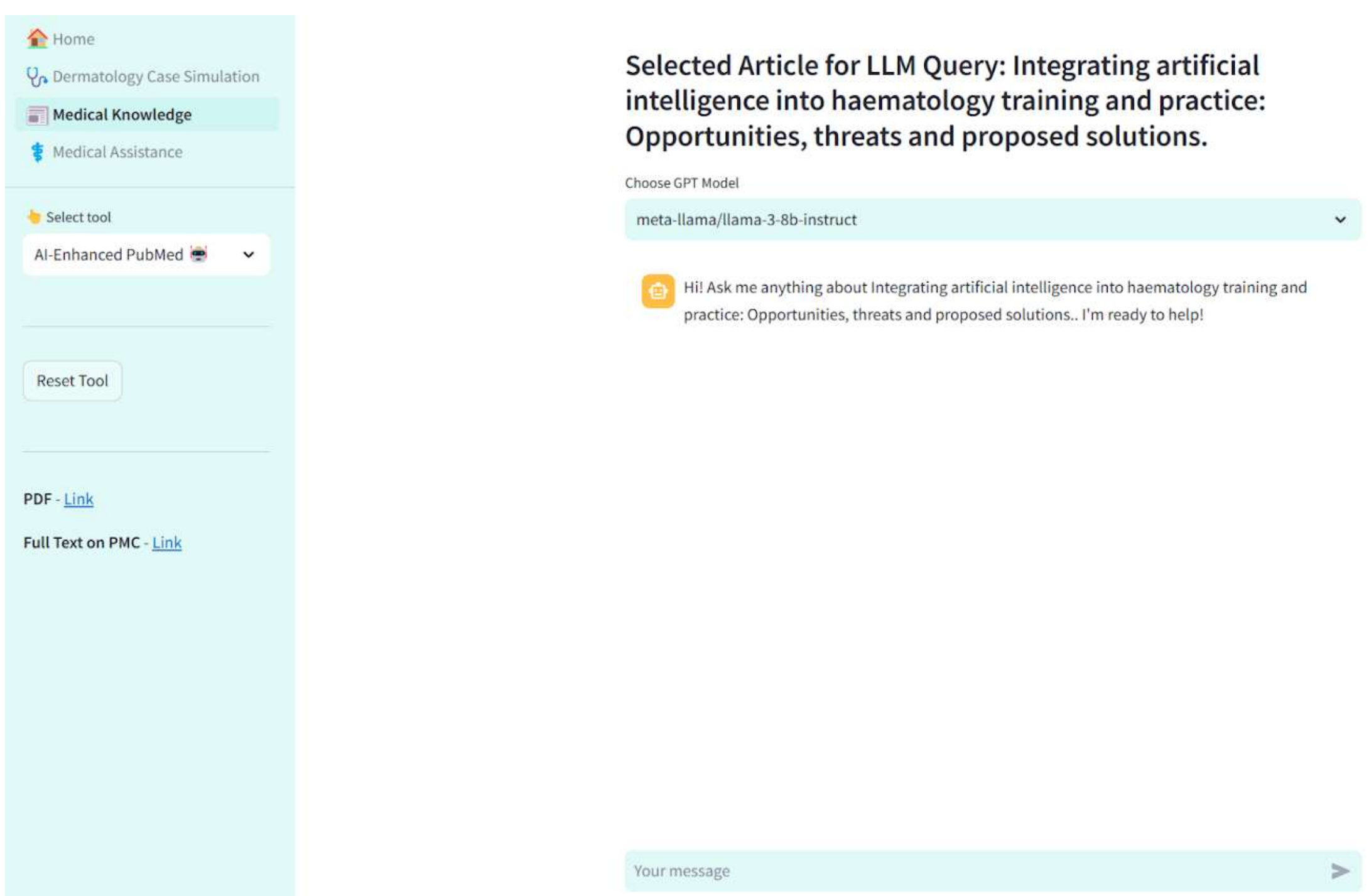

**Figure 13.** Chat interface in AI-enhanced PubMed Tool

## D. Usability Testing Survey

To evaluate the effectiveness and user satisfaction of *MediTools*, we conducted a usability testing survey among healthcare professionals and students. A total of 10 participants completed the survey.

### D.1 Demographics

The demographics section provides an overview of the respondents' roles, gender, age distribution, and years of experience in the medical field. **Table 1** below summarizes the roles among the respondents.

**Table 1**: Professional roles of survey participants

| Role | Frequency |
|---|---|
| Medical Student | 1 |
| Physician Assistant | 3 |
| Nurse | 3 |
| Resident | 1 |
| Attending Physician | 1 |
| Other | 1 |
| **Total** | **10** |

Our sample includes respondents from various professional roles. One respondent selected 'Other' and specified that they are a physician-scientist working in the industry. **Figure 14** presents a combined view of the gender, age, and years of experience distributions among the respondents. The majority, 60%, identify as female, while 40% identify as male. In terms of age, 70% of respondents are between 25-34 years old, and 30% are aged 18-24. Regarding healthcare experience, 50% of participants have 1-3 years of experience, 30% have 7-10 years, and 20% have less than 1 year.

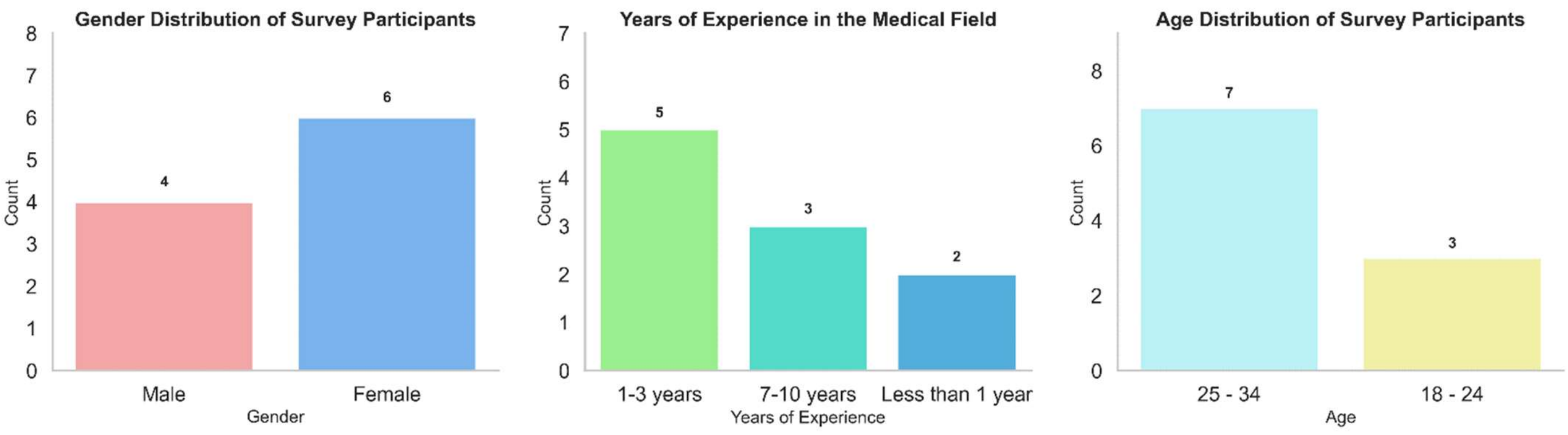

**Figure 14.** Gender, years of experience in healthcare, and age distributions among the respondents

## D2. Dermatology Case Simulation Tool Feedback

To evaluate the effectiveness of the dermatology case simulation tools, respondents rated the realism of the simulated patient interactions. As shown in **Figure 15**, 60% of participants rated the interactions as 'realistic,' while 40% rated them as 'very realistic,' indicating that the simulations closely resembled real patient encounters. Additionally, **Figure 16** presents participants' assessments of the LLMs' performance in understanding and responding accurately during the simulations, with 80% rating the performance as 'good' and 20% as 'excellent.' These findings suggest that the simulation tools provide a highly realistic and effective training environment.

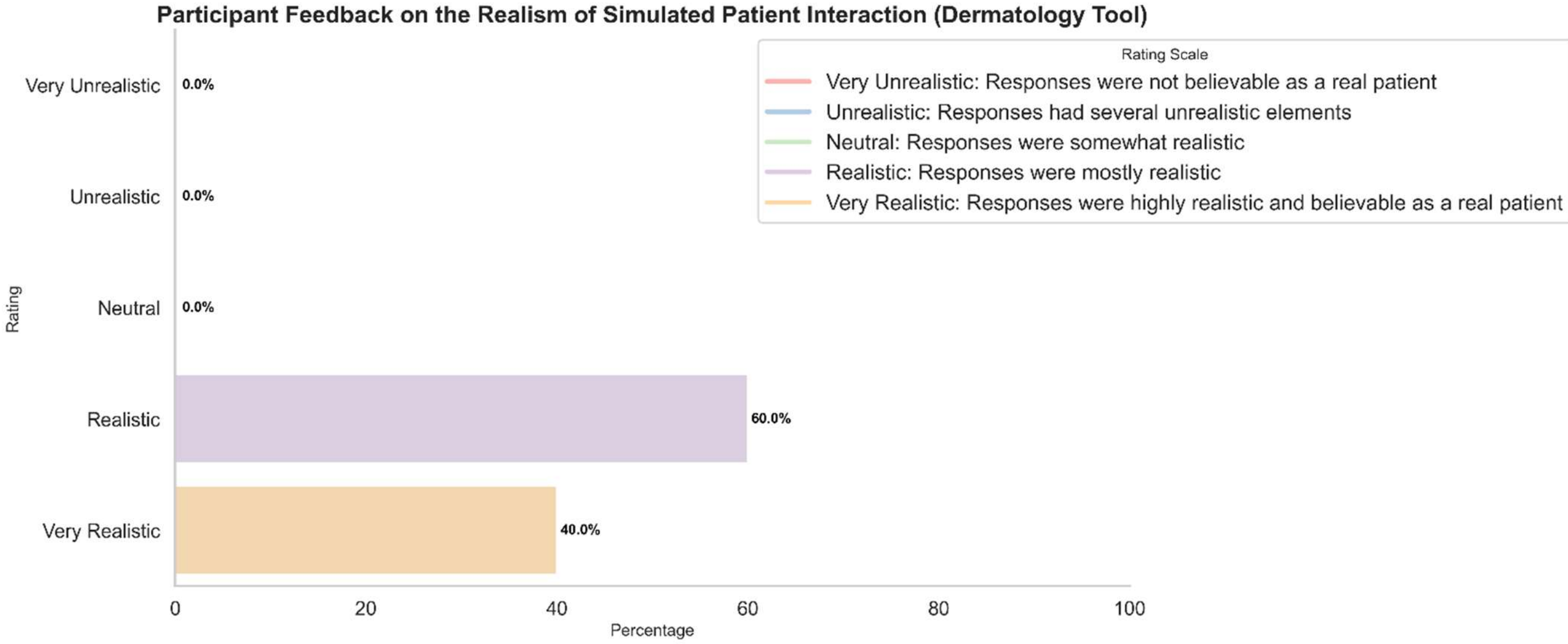

**Figure 15.** Participants feedback on realism of the simulated interaction. Question asked: "How realistic did you find the simulated patient interactions?"

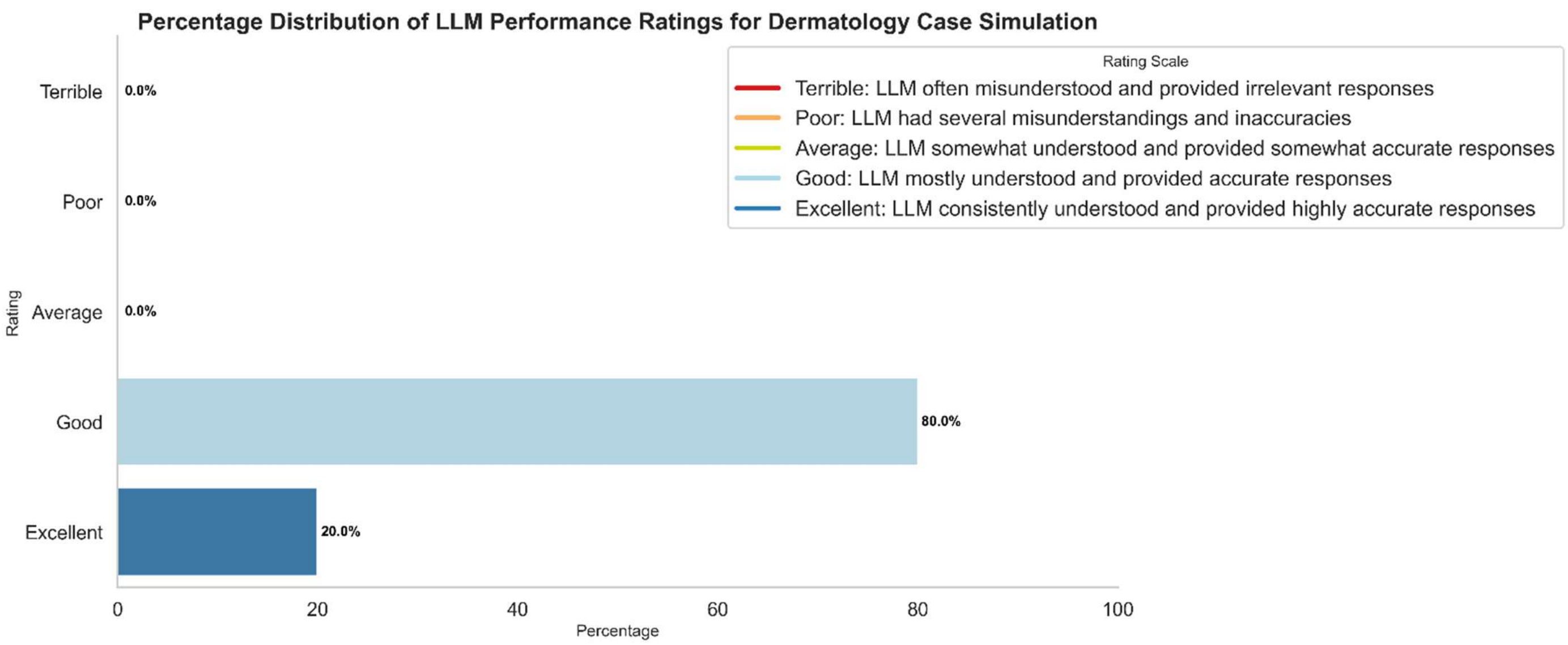

**Figure 16.** Participants' feedback on LLM performance. Question asked: "How would you rate the overall performance of the LLM in understanding and responding accurately?"

## D3. Google News Tool Feedback

**Figure 17** presents participants' assessments of the usefulness of LLM-generated summaries in the Google News tool. 20% percent of respondents rated the summaries as 'neutral,' indicating that they found the summaries moderately useful, containing some important details. However, 80% of participants rated the summaries as 'useful,' suggesting that they found the summaries to be mostly useful and inclusive of key details.

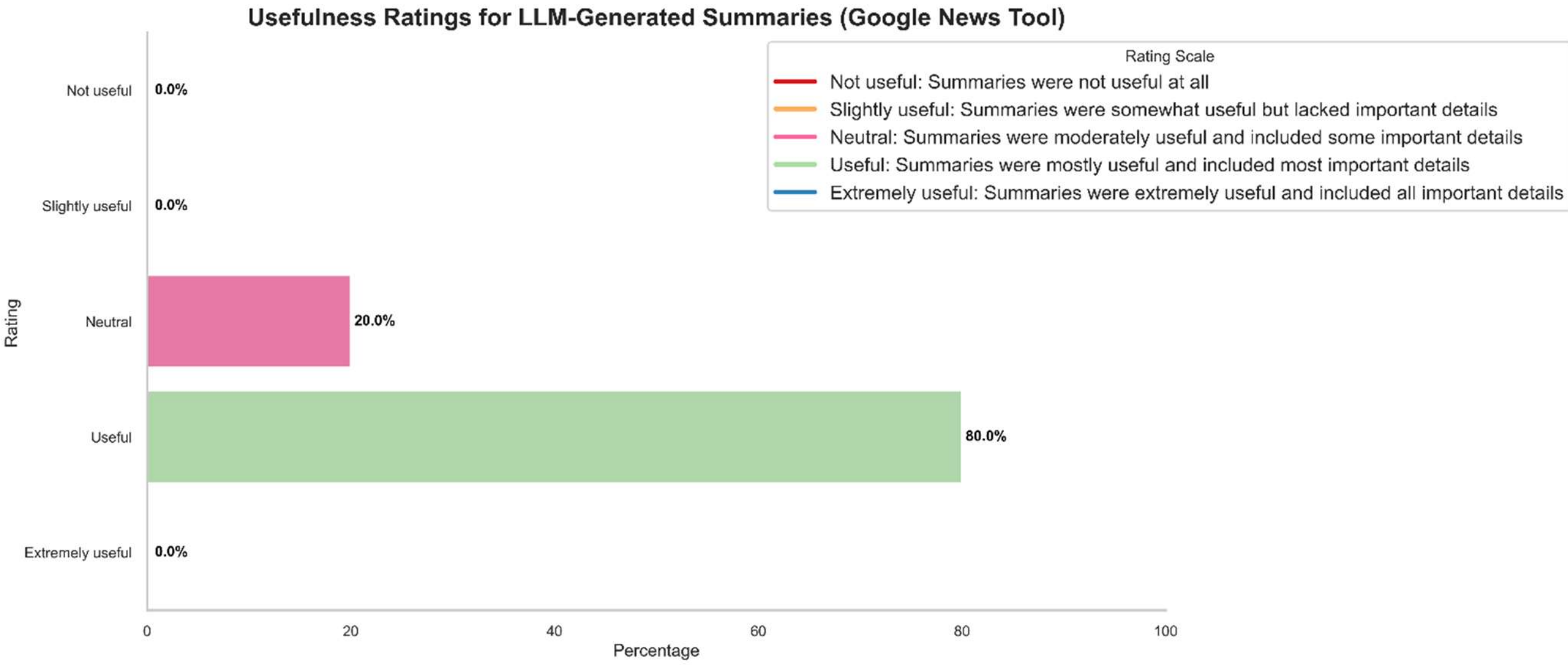

**Figure 17.** Respondents' feedback on usefulness of LLM generated summaries. Question asked: "How useful were the LLM-generated summaries of the articles?"

## D4. AI-Enhanced PubMed Feedback

**Figure 18** illustrates participants' ratings of the LLM's helpfulness in explaining research papers. The responses varied, with 10% finding it 'slightly helpful,' 20% rating it as 'neutral,' 50% considering it 'helpful,' and 20% rating it as 'very helpful.' For a detailed interpretation of these ratings, please refer to the legend in **Figure 18**.

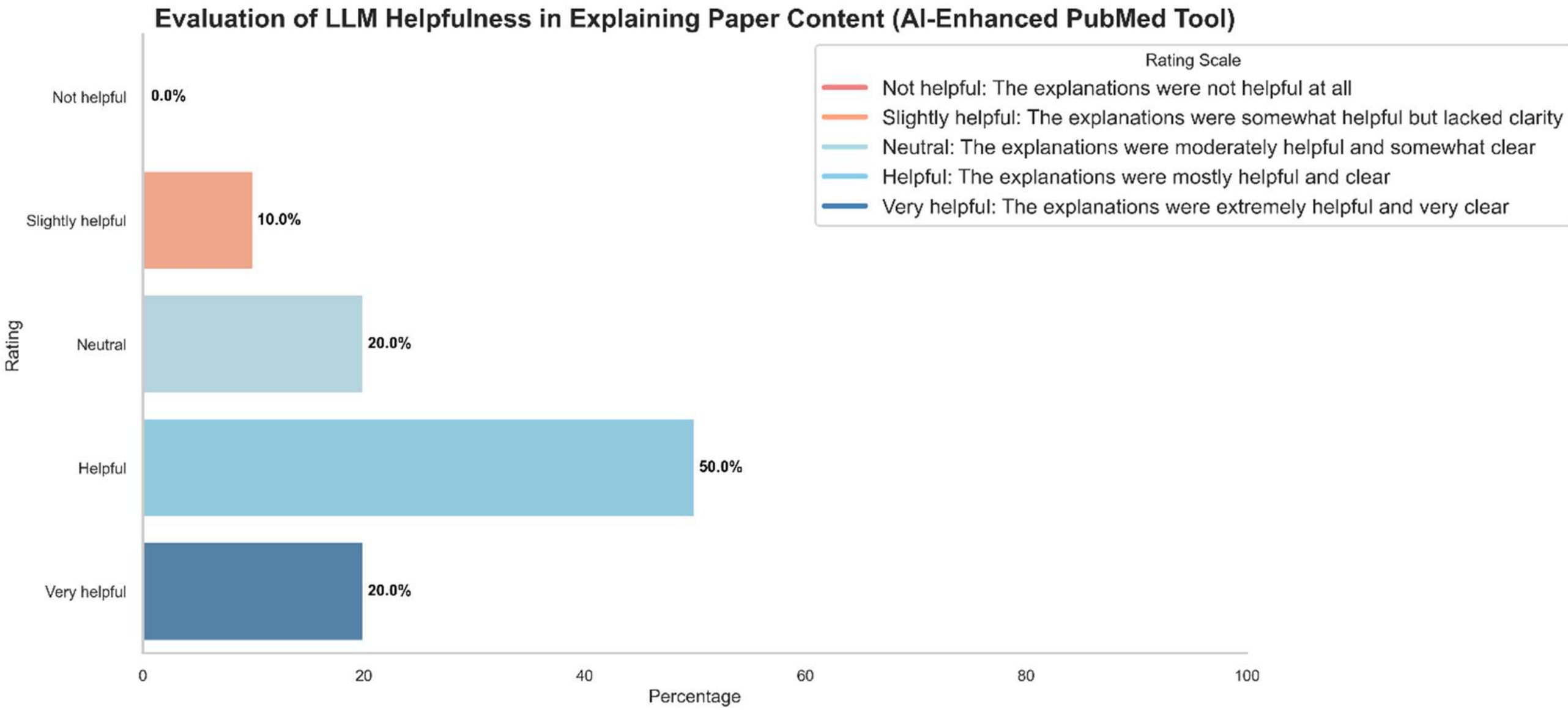

**Figure 18.** Participants' feedback on LLM helpfulness in explain content of research paper. Question asked: "How helpful was the LLM in explaining the content of the selected papers?"

## D5. MediTools Overall Feedback

To assess the overall usability of the *MediTools* application and its potential to enhance and modernize medical education, respondents were asked to evaluate whether the integration of AI and LLMs could lead to better learning outcomes. **Figure 19** displays the results, with 20% of respondents indicating 'probably yes' and 80% selecting 'definitely yes.' Additionally, **Table 2** presents responses to the question of whether participants would recommend *MediTools* to their peers. The results showed that 90% of respondents would recommend the application, while 10% chose 'maybe,' and none chose 'no.'

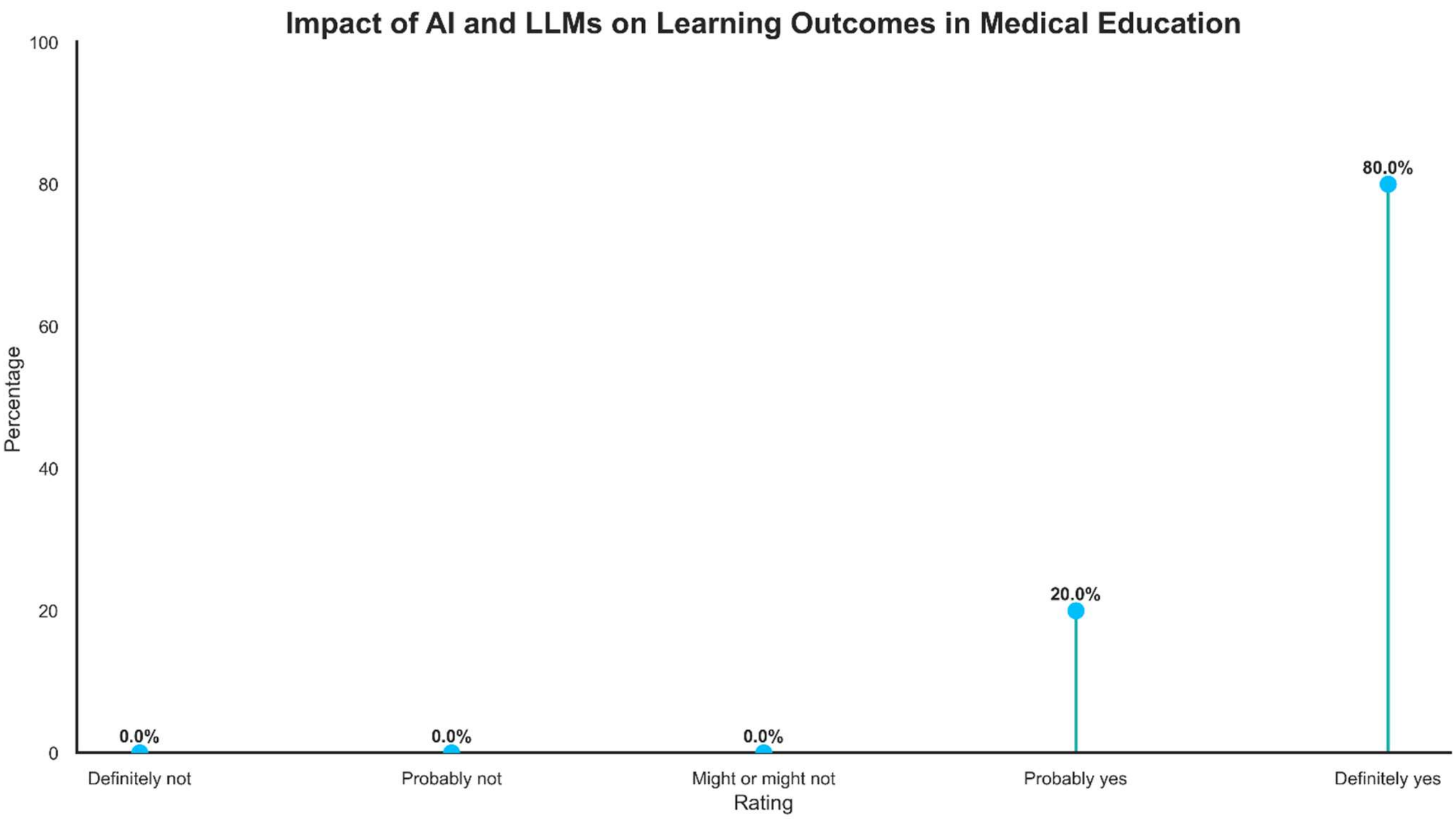

**Figure 19.** Respondents' perspective on the impact of AI and LLM on learning outcomes. Question asked: "Do you think the integration of AI and LLMs in medical education can lead to better learning outcomes?"

**Table 2.** Participants' Responses to the Question: "Would you Recommend MediTools to your Peers?"

| | *Frequency (Proportion of Total)* |
|---|---|
| *Yes* | 9 (90%) |
| *Maybe* | 1 (10%) |
| *No* | 0 (0%) |

## Discussion

The introduction of *MediTools* has showcased the potential of AI and LLMs in transforming medical education. The results from our study highlighted several aspects of this transformation. Leveraging LLMs can enhance the learning experience, our dermatology case simulation tool provides users with realistic patient interactions. Feedback from the collected survey indicated that these simulations were highly realistic and accurate. This suggests that LLMs can significantly enhance the learning experience by providing interactive and realistic scenarios for students to engage with. Additionally, our AI-enhanced PubMed Tool & Google News Tool demonstrated the potential to create more efficient ways to access and summarize medical literature and news with LLMs. Preliminary survey results indicate that users found the tools to be beneficial, with the dermatology case simulation tool providing a highly realistic and effective patient simulation. Feedback on the AI-enhanced PubMed and Google News tools highlights the potential of LLMs to aid users in understanding literature and gaining quick insights. Users expressed strong interest in *MediTools* and the integration of AI and LLMs in medical education, believing it could lead to improved learning outcomes. Notably, 90% of respondents indicated a willingness to recommend MediTools to their peers, reflecting a high level of satisfaction and perceived value in the tools developed. However, there is a need for continuous refinement of these tools to develop robust resources that facilitate continuous learning and provide easy access to medical information.

*MediTools* has shown us that the integration of LLMs into medical education represents a significant advancement in the field. Our beta testers and users have consistently praised the application's utility, noting that LLMs enhance the educational experience by providing immediate, relevant, and contextually appropriate responses. This form of technology is able to bridge the gap between theoretical knowledge and practical application, offering a dynamic and interactive learning environment for students and medical professionals. LLMs enable users to engage with realistic case scenarios, refine their diagnostic and communication skills, and stay updated with the latest medical research - faster and simpler than many current methods of achieving these tasks. More broadly, by automating administrative tasks and reducing the need for human actors or physical resources, LLMs make medical education more efficient and comprehensive, ultimately developing skilled and knowledgeable healthcare professionals.

While the results are promising, we encountered several challenges and limitations during this study. The development of this application was completed in three months with a fixed budget. As a prototype application, it relies on external sources and APIs to function properly. Despite the advanced capability of LLMs, there is still a risk of inaccuracies in the information generated. It is crucial that these models are evaluated to ensure the reliability and accuracy of AI-generated content. Ongoing monitoring and validation by human experts are necessary to maintain a high standard of quality. Also, the integration of these tools requires robust technical infrastructure and resources, institutions must be prepared to invest in these areas to fully realize the benefits of AI-based educational tools. The development of *MediTools* is a step forward in the integration of AI and LLMs into medical education. Future research and development should focus on enhancing existing tools and developing new applications to cover a range of use cases and training scenarios. There is a need for more long-term studies that assess the impact of AI and LLM-based tools on learning outcomes, retention of knowledge, and clinical performance. Continuous collaboration between stakeholders is crucial to creating comprehensive and effective AI-driven educational ecosystems.

The Integration of AI and LLMs into medical education, as demonstrated by our prototype application *MediTools*, holds significant promise to enhance the learning experience. As we continue to refine these technologies and expand their application, AI and LLMs are positioned to revolutionize and modernize medical education, ultimately leading to better learning outcomes, better prepared healthcare professionals, and improved patient care.

## References

Abd-alrazaq1, A., AlSaad1, R., Alhuwail3, D., Ahmed1, A., Healy4, P. M., Latifi4, S., Aziz1, S., Damseh5, R., Alrazak6, S. A., Sheikh1, J., Health, 1AI Center for Precision, & Abd-alrazaq, C. A. (2023). *Large language models in medical education: Opportunities, challenges, and future directions*. JMIR Medical Education. https://doi.org/10.2196/48291

Furfaro, D., Celi, L. A., & Schwartzstein, R. M. (2024). Artificial Intelligence in Medical Education: A Long Way to GO. *CHEST*, *165*(4), 771–774. https://doi.org/10.1016/j.chest.2023.11.028

Li, Y., & Li, J. (2023). Generative Artificial Intelligence in Medical Education: Way to solve the problems. *Postgraduate Medical Journal*, *100*(1181), 203–204. https://doi.org/10.1093/postmj/qgad116

Nagi, F., Salih, R., Alzubaidi, M., Shah, H., Alam, T., Shah, Z., & Househ, M. (2023). Applications of artificial intelligence (AI) in Medical Education: A scoping review. *Studies in Health Technology and Informatics*. https://doi.org/10.3233/shti230581